\documentclass{aa}  

\usepackage{graphicx}
\usepackage{txfonts}
\usepackage{xcolor} 

\usepackage{soul} 

\begin{document}

   \title{Slow and sausage loop mode excitation due to local and global spontaneous perturbations}


   \author{H. Capettini\inst{1,2}
          \and
          M. Cécere\inst{1,3}
           \and
          A. Costa\inst{1}
          \and
          G. Krause\inst{4,5}
          \and
          O. Reula\inst{6,2}
          }

   \institute{Instituto de Astronom\'{\i}a Te\'orica y Experimental, CONICET-UNC, C\'ordoba, Argentina.
         \and
             Facultad de Matem\'atica, Astronom\'{\i}a, F\'{\i}sica y Computaci\'on, UNC, C\'ordoba, Argentina.
         \and
             Observatorio Astron\'omico de C\'ordoba, UNC, C\'ordoba, Argentina.
        \and
             Instituto de Estudios Avanzados en Ingenier\'{\i}a y Tecnolog\'{\i}a, CONICET, C\'ordoba, Argentina.
        \and
             Facultad de Ciencias Exactas, F\'{\i}sicas y Naturales, UNC, C\'ordoba, Argentina.
        \and
             Instituto de F\'{\i}sica Enrique Gaviola, CONICET-UNC, C\'ordoba, Argentina.\\
             \email{mariana.cecere@unc.edu.ar}
             }

   \date{Received July 24, 2020; accepted October 21, 2020}

\begin{abstract}
      {}
   {We analyse the capability of 
   different type of perturbations -associated with usual
environment energy fluctuations of the solar corona- to excite slow and sausage modes in solar flaring
loops.}
   {We perform  numerical simulations of the MHD ideal equations considering straight plasma magnetic tubes subject to local and global energy depositions.}
    {We find that  local loop energy depositions of typical microflares [$\sim$($10^{27}$-
$10^{30}$) erg] are prone to drive slow shock waves that induce  slow mode patterns. The slow mode
features are obtained for every tested local energy deposition inside the loop.  
Meanwhile, to obtain an
observable sausage mode pattern a global perturbation, capable to modify instantaneously the internal
loop temperature, is required, i.e. the characteristic conductive heating time must be much smaller than the radiative cooling one. 
Experiments carried out by varying parameter $\beta$ show us that the excitation of sausage modes does not depend significantly on the value of this parameter but on the global or local character of the energy source.}
  {}
   \keywords{Sun: oscillations -- Magnetohydrodynamics (MHD) -- Shock waves
               }
   \maketitle
   \end{abstract}
%

\section{Introduction}

The coronal seismology is a later branch of the solar physics that combines the measurement of temporal and spatial signatures of magnetohydrodynamics (MHD) waves and oscillations together with  their theoretical modelling  in different magnetic structures to infer coronal mean plasma properties. 
In the particular case of the coronal loops, the fast sausage mode and the standing slow mode are the most  commonly studied due to their compressibility which makes them susceptible to be observed \citep{2001A&A...372L..53N,2005ApJ...624L..57A,2007A&A...463..333A,2011ApJ...727L..32V}.

Sausage modes modelled as axisymmetric perturbations of magnetic cylinders
 are known to be compressible MHD fast modes that perturb 
the plasma inside the loop in the radial direction causing a symmetric contraction and a widening of the tube without distortion of its axis \citep{1983SoPh...88..179E}.
These modes have characteristic periods of $\sim[5\!-\!10]\,\text{s}$ and decaying times of few periods and they are generally found associated with  flaring loops in active solar regions  \cite[e.g.,][]{1989SvAL...15...66Z,2001ApJ...562L.103A,2003A&A...412L...7N,2004AstL...30..480S,2008MNRAS.388.1899S,2011ApJ...740...90V}. 
In fact, \citet{2004ApJ...600..458A} showed that the existence of trapped sausage modes, prescribed by their cutoff, requires the host loop to have a high electron density which can only occur in flaring loops. 
The work by \cite{2012ApJ...761..134N} has important seismological implications for the diagnosis of plasmas in flaring loops using sausage oscillations. They showed that, in the linear regime, an initial perturbation results in
either a leaky or a trapped regime of sausage oscillation, depending upon whether the longitudinal wavenumber is smaller
or greater than a cutoff value, respectively.
Also, if the wave damping caused by the leakage is not 
high enough  the oscillations may be
detectable for several cycles. Specially in dense and thick flaring loops, where the lifetime of high radial sausage harmonics could be sufficiently long, these leaky oscillations could be responsible for observed quasi-periodic pulsations (QPP) \citep{2005SSRv..121..115N, 2007A&A...461.114P,2020ApJ...893...62L}.

\citet{1989SvAL...15...66Z} proposed that a modulation of hard X-ray and white-light emission from a loop footpoint can occur due to the change in the loop radius and the consequent change of the mirror ratio, causing the periodic precipitation of non-thermal electrons at flaring loops.
Moreover, from QPP in  hard X-ray and microwave time profiles, \citet{2001ApJ...562L.103A} proposed that a modulation can be produced by the variations of macroscopic magnetic structures, for instance, oscillations of coronal loops.
Analysing a flaring loop in microwaves, \citet{2008A&A...487.1147I} showed that the whole loop oscillates with the same period and almost in phase producing either a MHD sausage mode or a periodic regime of magnetic reconnection.
Also, EUV and soft X-ray band periodic variations of the thermal emission intensity were associated with density perturbations and Doppler shift of the emission lines leading to the detection of sausage modes \citep{2016ApJ...823L..16T}.
\cite{2010ApJ...712L.111J} and \cite{2018ApJ...859..154N} presented interesting observational examples of the excitation of MHD oscillations in a coronal loop by a microflare. These observations are  in line with our motivation to simulate the  excitation of sausage modes by deposition of energy pulses of short duration (we call them impulsive excitations because the duration of the pulse is a small fraction of the period of oscillation). 

Slow magnetoacoustic modes are more commonly excited widespread throughout the corona. In loops, they are generally thought off as standing longitudinal modes with periods of the order of several minutes and with decaying times of few periods \citep[e.g.][]{2000A&A...362.1151N, 2003A&A...406.1105W,2005A&A...435..753W,2019ApJ...874L...1N}.
Numerous  observations have reveled that these oscillations can be triggered by hot impulsive flares or smaller brightening located somewhere in the loop, e.g., close to a loop footpoint  or near the apex \citep{2002A&A...381..311D, 2002SoPh..209...89D,2004SoPh..222..229C,2006ApJ...639..484M}.
Numerical simulations emulate the impulsive flares by pulses of energy depositions that trigger numerical features resembling slow magnetosonic standing waves \citep[e.g.,][]{2005A&A...436..701S,2007ApJ...668L..83S,2007A&A...467..311O,2009MNRAS.400.1821F,2009A&A...495..313O}.

To study the capability of producing sausage modes we study different energy release scenarios resembling the action of microflares. To that end, we analyse the action of, firstly, an impulsive local deposition and, secondly, impulsive global energy depositions capable to  excite a pattern of coupled modes. With local deposition of energy we refer to an energy perturbation that is initially circumscribed in a small region inside the loop and produces a disequilibrium in the loop interior.
Consistently, with global energy depositions we refer to an energy perturbation that initially covers the whole loop and produces its disequilibrium with the external corona. 
The quantitative description of these models is given below.

\section{The model}
The basic model starts with the ideal MHD equations that describe the macroscopic behaviour of a compressible ideal fully ionized plasma. The ideal MHD equations   in its conservative form and in CGS units are written as:
\begin{equation}\label{e:cont}
\frac{\partial\rho}{\partial t}+\nabla\cdot(\rho\mathbf{v})=0 \, ,
\end{equation}
\begin{equation}\label{e:euler}
\frac{\partial (\rho \mathbf{v})}{\partial t} + \nabla \cdot \left(\rho \mathbf{v} \mathbf{v} - \frac{1}{4\pi} \mathbf{B}\mathbf{B} \right) + \nabla p + \nabla\left( \frac{B^2}{8\pi}\right)   = \mathbf{0} \, ,
\end{equation}
\begin{equation}\label{e:consE}
\frac{\partial E}{\partial t} + \nabla \cdot \left[\left(E + p + \frac{B^2}{8\pi}\right)\mathbf{v} -\frac{1}{4\pi} \left(\mathbf{v\cdot B}\right)\mathbf{B}\right] = 0 \, ,
\end{equation}
 \begin{equation}\label{e:induccion}
\frac{\partial \mathbf{B}}{\partial t} + \mathbf{\nabla \cdot} \left(\mathbf{v} \mathbf{B} - \mathbf{B} \mathbf{v} \right) = \mathbf{0} \, ,
\end{equation}
where no gravitational terms are considered, $\rho$ indicates the plasma density, $p$ the thermal pressure, $\mathbf{v}$ the velocity, $\mathbf{B}$ the magnetic field and $E$ is the total energy (per unit volume) given by  
\begin{equation*}
    E = \rho \epsilon + \frac{1}{2} \rho v^2 + \frac{B^2}{8\pi},
\end{equation*}
where $\epsilon$ is the internal energy.

In addition to the MHD equations, the divergence-free condition of the magnetic field, i.e.
\begin{equation}\label{e:divB}
 \mathbf{\nabla\cdot} \mathbf{B} = 0\, ,
\end{equation}
must be fulfilled.

To complete the set of MHD equations a closure relation among the thermodynamic variables must be imposed. We assume a calorically perfect gas for which $p = 2\rho k_B T/m_i = (\gamma - 1) \rho \epsilon$, where $k_B$ is the Boltzmann constant, $T$ the plasma temperature, $m_i$ the proton mass (assuming that the plasma is fully ionized hydrogen), and $\gamma = 5/3$ is the specific heat relation.

\subsection{Equilibrium configuration}
\begin{figure}
 \centering
\includegraphics[width=5cm]{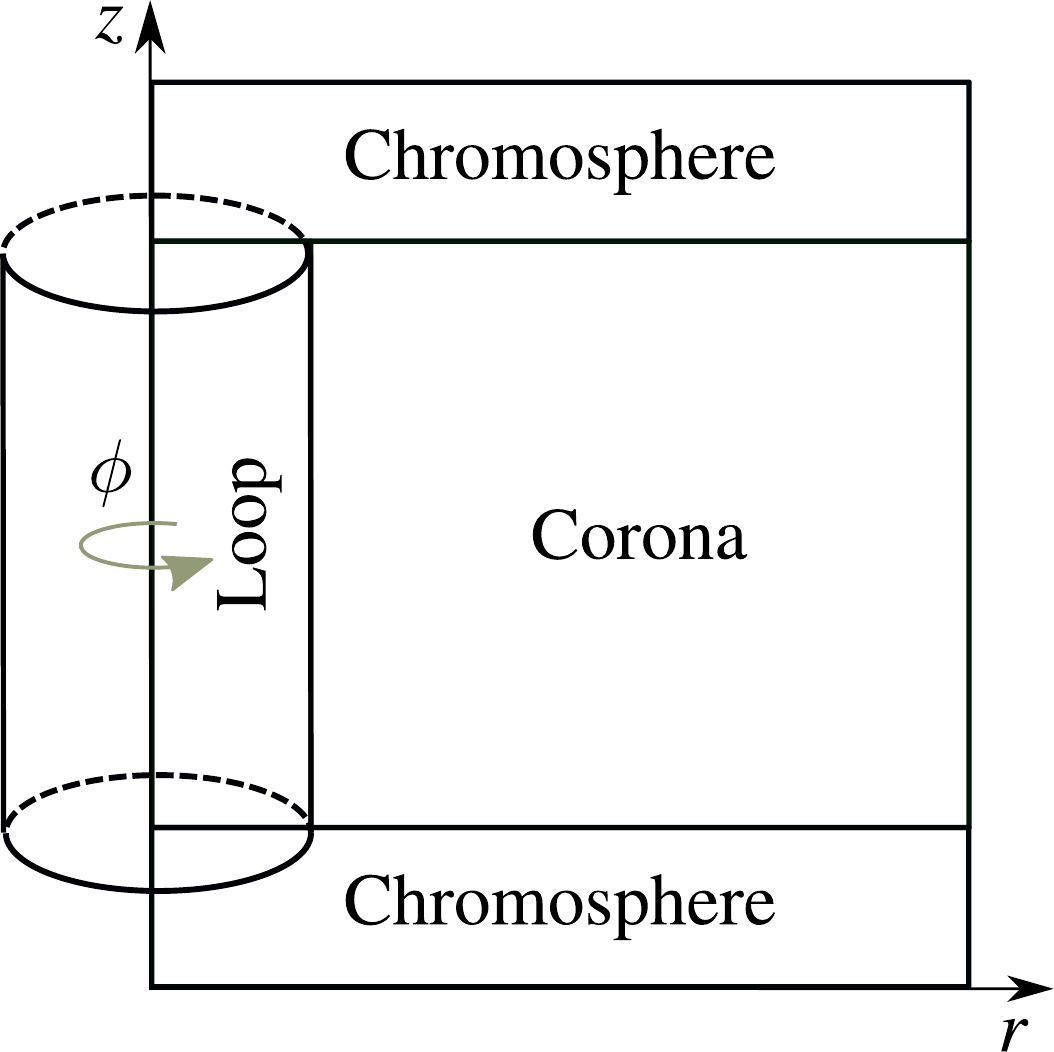}
 \caption{Scheme of the coronal loop, corona and chromosphere in the equilibrium configuration.}
 \label{fig:modelo}
\end{figure}

The equilibrium configuration consists of a straight cylinder surrounded by the solar corona and a chromosphere at its ends. We impose an axial symmetry around $r = 0$  and we use a 2.5D simulation (in cylindrical coordinates), allowing to save  computational time,  that could be raised to a 3D simulation removing the azimuthal symmetry (see  Fig.~\ref{fig:modelo}). To achieve the overall equilibrium setup the different initial plasma parameter values are given in Table~\ref{tab:table1}.

\begin{figure}
 \centering
  \includegraphics[width=9cm]{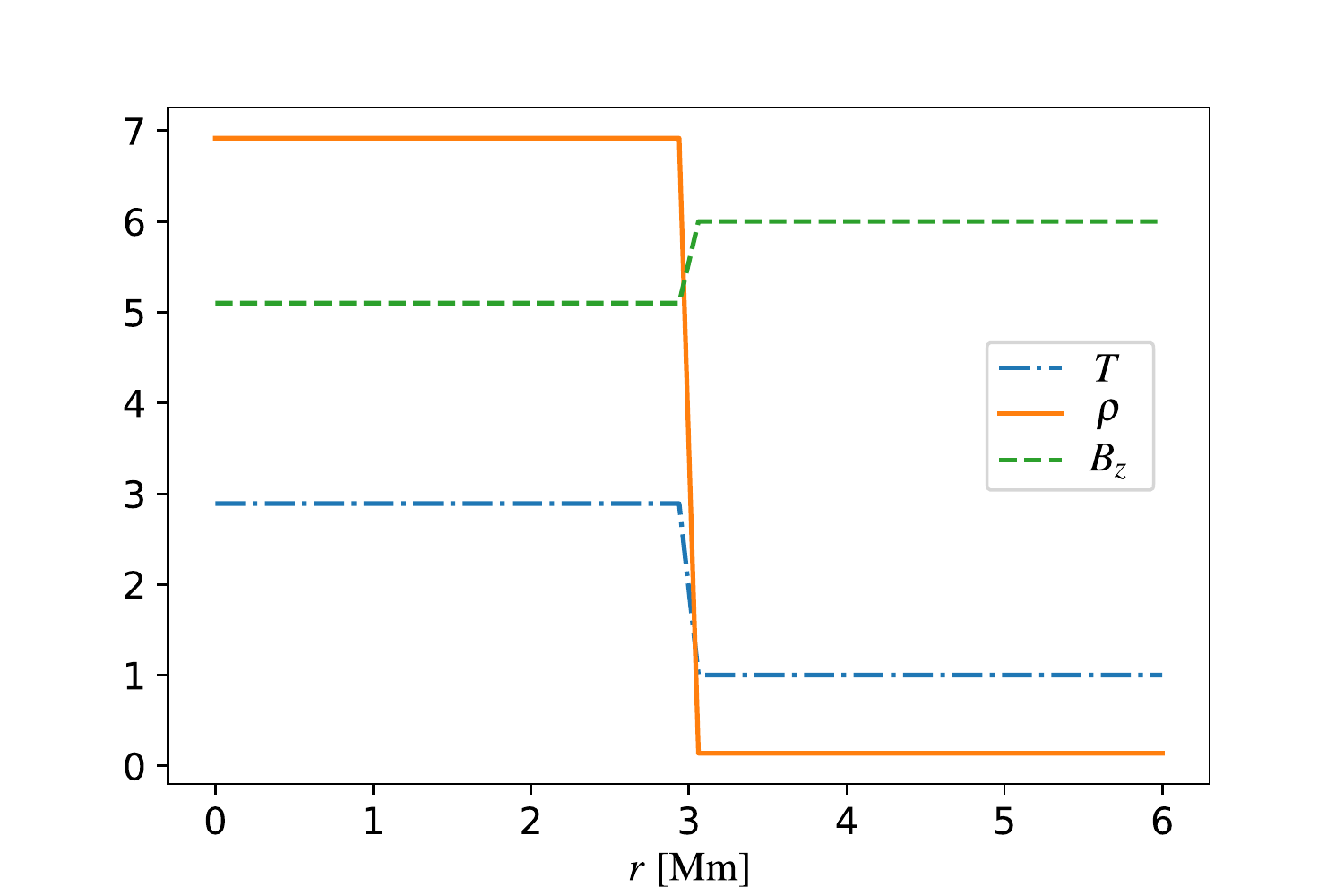}
 \caption{Plot of the initial parameters as a function of the transverse coordinate. The solid line is the density (g cm$^{-3}$), the dashed line is the magnetic field $B_z$ (in G), and the dot-dashed line is the temperature (MK).  For clarity the density and the magnetic field were divided by a factor $1.2\times10^{-14}$ and $10$, respectively. }
 \label{fig:equilibrio}
\end{figure}

In the model the magnetic field is taken to be in the $z$-direction, $\mathbf{B} = (0,0,B_z)$
and the environment is motionless ($\mathbf{v} = \mathbf{0}$). 
Under these conditions and using Eq.~\eqref{e:euler} it can be seen that the equilibrium is obtained if 
\begin{equation*}
 p_l + \frac{B_l^2}{8 \pi} = p_0 + \frac{B_0^2}{8 \pi}\, ,
\end{equation*}
where the sub-index $l$ states for loop parameters and the sub-index $0$ for coronal parameters.
The initial state variables profile can be seen in  Fig.~\ref{fig:equilibrio}. The $\beta$ parameter is defined by $8\pi p/B^2$.
In flaring coronal loops, oscillations can occur  either with $\beta > 1$ \citep[e.g.,][]{2001ApJ...557..326S} or with $\beta < 1$ as reported by \citet{2003A&A...412L...7N}, which is the case that we take as reference in this work.
In order              to control the value of $\beta < 1$  inside the loop, appropriate temperature, magnetic fields and density are chosen. Under this assumption,  the simulated plasma structures are hotter and denser than the surrounding medium, which is consistent with the known properties of flaring coronal loops.

\begin{table*}[h]
 \centering
 \begin{tabular}{|c|c|c|c|c|}
  \hline
  & Corona     & Loop     & Chromosphere/Corona & Chromosphere/Loop  \\
  \hline
  \hline
  Density [g cm$^{-3}$] & $1.66\times 10^{-15}$ & $8.33\times 10^{-14}$ & $1.66\times 10^{-13}$  & $8.33\times 10^{-12}$ \\
  \hline
  Temperature [MK] & $1.000$ & $2.890$ & $0.010$ & $0.0289$ \\
  \hline
  $B_z$  [G] & $60$ & $51$ & $60$ & $51$ \\
  \hline
  $\beta$ & $0.002$ & $0.38$ & $0.002$ & $0.38$ \\
  
  \hline
 \end{tabular}
 \caption{System initial equilibrium values. }
 \label{tab:table1}
\end{table*}
    
\subsection{Perturbations}
\label{perturbations}

To analyse the modes that can be excited in the loop structure we perturb  the initial equilibrium state with an axisymmetric perturbation. Firstly, we  investigate the conditions that would lead to the excitation of a sausage mode. Thus, we start perturbing the equilibrium with a localised deposition of energy injected at different loop heights,  emulating the energy deposition of  typical microflares: $1.5\times 10^{27}\,\text{erg}$ and $3.5\times10^{28}\,\text{erg}$.  For this purpose, we confined the energy deposition to a sphere of radius $R=1\,$Mm localised in the center of the loop ($r=0\,\text{Mm}, z=17.5\,\text{Mm}$) with radius  $R_l=3\,$Mm and length $L=25\,$Mm \citep{2003A&A...412L...7N}. These scales determine the effectiveness of the linear excitation of different harmonics, as it has been shown in \cite{2005SSRv..121..115N,2007A&A...461.114P} and \cite{2020ApJ...893...62L}.

We  also modify the equilibrium yielding a global perturbation to the system. In this case, the size of the perturbation is the size of the loop. 
This type of perturbation can be attributed to an energy injection at the loop bases occurring in a flaring region.
When a flare occurs, a large amount of energy is released in few seconds. The heat is conducted along the whole loop and -due to the anisotropic heat conduction rate that inhibits conduction across the magnetic field lines-   the loop absorbs energy beyond its environment energy content leading to an imbalance with the surrounding medium.
When the conduction characteristic time is larger than the characteristic radiative cooling time we assume that the excess of heat at the loop interior with respect to its environment can be interpreted as a global perturbation or a global deposition of energy. 

The \emph{thermal conduction time} is
\begin{equation}
 \tau_{\mathrm{cond}} = \frac{3nk_{\mathrm{B}}L^2}{\kappa_0 T^{5/2}} \, ,
\end{equation}
and the \emph{radiative cooling time} is
\begin{equation}
 \tau_{\mathrm{rad}} = \frac{3k_{\mathrm{B}}T }{n\Lambda(T)}\,,
\end{equation}
where $n$ is the number density, $L$ is a characteristic length, $k_0 \propto 10^{-6}\,\text{erg}\,\text{K}^{-7/2}\,\text{cm}^{-1}\,\text{s}^{-1}$ is the
heat conduction coefficient along the magnetic field and $\Lambda(T)$ is the radiative loss function
($10^{-23}\,\text{erg}\,\text{cm}^{-3}\,\text{s}^{-1}$ for $T \sim 10\,\text{MK}$) \citep{2005psci.book.....A}.

Note that, for a flaring region with
$T = 40\,\text{MK}$, a short loop length of $L = 25\,\text{Mm}$  with number density  $n = 5\times 10^{10}\,\text{cm}^{-3}$ \citep{2003A&A...412L...7N}, the conductive and radiative characteristic times are  $\tau_{\text{cond}} = 20\,$s, and
$ \tau_{\text{rad}} = 20000\,\text{s}$, respectively.
Thus,  $\tau_{\text{cond}} \ll \tau_{\text{rad}}$ and the excess of heat content cannot be cooled down by radiation in times of the conductive time order. Therefore,  a global perturbation occurs associated to a short and dense loop in a flaring region, where sausage loops are generally observed \citep{2004ApJ...600..458A}.  

On the other hand, in a quiet sun region, where the temperature is approximately $T = 2\,\text{MK}$, and for the loop parameters $L = 25\,\text{Mm}$ and $n = 5\times 10^{10}\,\text{cm}^{-3}$, the conductive and radiative characteristic times are $\tau_{\text{cond}} = 200000\,\text{s}$, $\tau_{\text{rad}} = 40\,\text{s}$, respectively.
Thus, ${\tau_{\text{rad}} \ll \tau_{\text{cond}}}$; 
given a typical energy perturbation, the heat content cannot be retained along the whole loop because the radiation proceeds to cool it down. In this case, a local perturbation is responsible for the oscillating pattern. 

We analyse the signatures that these perturbations produce when  tube modes are excited and describe their characteristic Fourier transforms (FT). A slow, mostly longitudinal, tube  mode has the characteristic that the phase where the   axial velocity is increase ($v_z$) is correlated with an increase of the axial magnetic field component ($B_z$) and a decrease of  the density, and vice versa. On the other hand, the phase where the sausage mode symmetrically contracts the tube  ($R$ is decreasing) is correlated with an increase of  the axial magnetic field component ($B_z$; the field lines are closer)  and an increase of the density, and vice versa.
Hence, the Fourier transform signature of a slow tube mode is associated with  a density peak at the slow frequency, correspondingly   with a peak in the axial velocity ($v_z$) and the magnetic field ($B_z$) components, at the same frequency.  The FT signature of the sausage mode is associated with  a density peak at the fast frequency correspondingly   with a peak in the radial velocity ($v_r$) and the magnetic field ($B_z$) components, at the same frequency.
 
In most observations, the elementary bursts seem to arise from a single flaring loop than from several of them, resulting not from a sequence of energy-release processes but rather from a single deposition in a determined one \citep{1989SvAL...15...66Z}. But also, the perturbation could   take some time, i.e., the driver could be time-dependent, which could  also be important for the excitation of MHD modes \citep{2019A&A...624L...4G}. In this work, we choose to trigger the numerical experiments with an instantaneous deposition of energy. 

\subsection{Numerical code}
In order to evaluate the plasma behaviour, the MHD Eqs.~\eqref{e:cont}--\eqref{e:induccion} are numerically solved in  Cylindrical grid of co-located finite volumes.
We perform 2.5D simulations considering axial symmetry of the model with the $z$-direction parallel to the axis of the magnetic loop, neglecting the loop curvature. The simulations  are carried out using the FLASH Code \citep{2000ApJS..131..273F}, an open-source publicly available suite of high-performance simulation tools developed at the Center for Astrophysical Thermonuclear Flashes (Flash Center) of the University of Chicago.
This code, currently in its fourth version, uses the finite volume method with Godunov-type schemes to solve the high energy compressible MHD equations on regular grids. For our simulations we choose the unsplit staggered mesh (USM) solver available in FLASH,  which  uses a second-order directionally unsplit scheme with a MUSCL-type reconstruction. This solver implements a more consistent treatment of the magnetic field, since its formulation is based on the constrained transport method and the corner transport upwind method, which avoids the generation of non-physical magnetic field divergence \citep{2009JCoPh.228..952L}.
To solve the interface Riemann problems we set the HLLD solver among the available options.

Cylindrical 2D square grids are used to represent the physical domain  of $[0,35]\,\text{Mm}\times [0,35]\,\text{Mm}$ with a discretisation of $350 \times 350$ cells, obtaining a resolution of $\sim [0.1 \times 0.1]\,\text{Mm}^2$.
Boundary conditions are set as follows. At the right lateral end outflow conditions (zero-gradient) are applied for all the variables allowing waves to leave the domain without reflection.
At the left lateral ($r = 0\,\text{Mm}$), a mirroring boundary condition is set for the vector fields in order to represent the cylindrical symmetry in the $\phi$ variable. At the bottom ($z = 0\,\text{Mm}$) and at the top ($z = 35\,\text{Mm}$) of the simulation we implement line tied boundary conditions to emulate the behaviour  of the dense  chromosphere, which extends  $5\,\text{Mm}$ from each end.

\section{Results and discussion}
 In this section we present the results for the two considered scenarios,  the localised energy depositions and the global instantaneous energy deposition, to evaluate what kind of perturbations are feasible to develop at each of them.
After that, we analyse the influence of the plasma features in the dynamic of the oscillations.

\subsection{Localised energy depositions}
\label{sec:local}
An instantaneous pulse of $1.5\times10^{27}\,\text{erg}$ in a sphere of radius $R = 1\,\text{Mm}$, temperature  $8\,\text{MK}$ and $n = 10^{10}\,\text{cm}^{-3}$, resembles the energy deposition released by a typical microflare \citep{2005psci.book.....A}.
This perturbation produces a spherical shock wave which in the evolution becomes highly collimated in the $z$-direction by the action of the loop magnetic field.
Along the  $z$-direction the magnetic field plays the role of being the waveguide of a fundamentally hydrodynamic shock \citep{2009MNRAS.400.1821F}. A pair of opposite slow  shock fronts develop and, when the chromosphere is reached, they are mainly reflected inside the loop. 

Also, a fast shock wave front pops out in the radial direction reaching the coronal environment through the loop's boundary. \footnote{The loop parameters are such, that linear perturbations would lead to  trapped modes \citep{2003A&A...412L...7N} (see equation 5 of that paper), however, as the perturbations we are working with are nonlinear, the trapped mode condition is not valid and the energy can leak through the loop boundary.}

In the following, we choose the apex (i.e.,  the middle point in the cylinder) and the base of the loop as the locations for the local energy depositions to perform  numerical experiments that emulate the action of typical microflare events.

\subsubsection{At the apex}
\begin{figure}
  \centering
  \includegraphics[width=9cm]{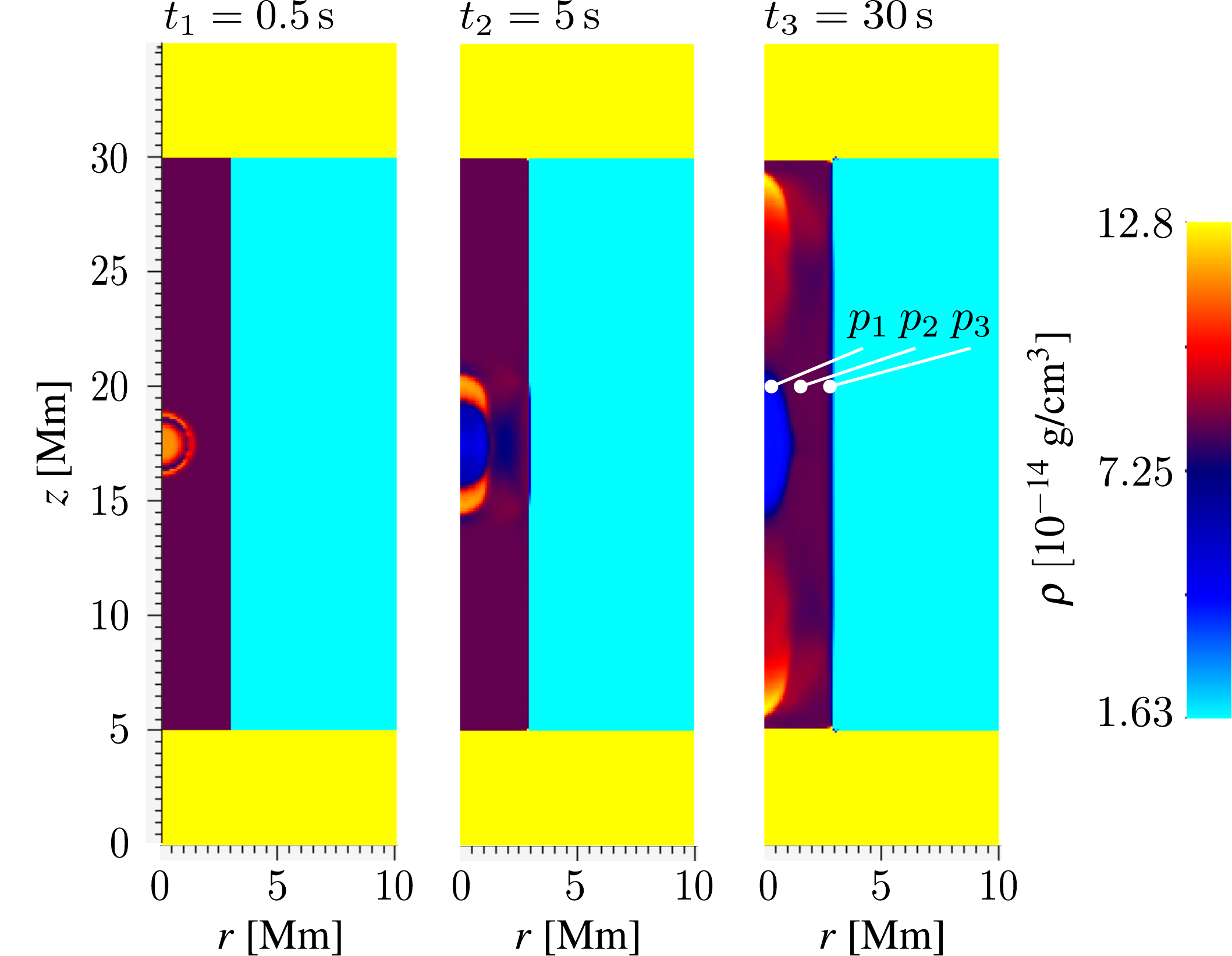}
  \caption{Density evolution  at different times. The minimum and maximum values chosen highlight the shock wave fronts. The points $p_1$, $p_2$ and $p_3$ mark the positions where the perturbation measure is performed.   }
        \label{fig:shock_loop}
 \end{figure}

In this case, the pulse is located at the center of the loop $r = 0\,\text{Mm}$, $z = 17.5\,\text{Mm}$.
Figure~\ref{fig:shock_loop} shows a snapshot of the numerical domain and the evolution for three different times $t_1 = 0.5\,\text{s}$, $t_2 = 5\,\text{s}$, $t_3 = 30\,\text{s}$;  note the evolution of the shock front, initiated as a sphere which becomes collimated in the $z$-direction due to the magnetic field orientation.
We choose three loop positions to measure the space and time action of the perturbations on the plasma  properties: $r = (0.2, 1.5, 2.8)\,\text{Mm}$ with $z = 20\,\text{Mm}$, indicated as $p_1$, $p_2$ and $p_3$, respectively, in the right panel of the figure (see $t_3$). 

\begin{figure}[h!]
  \centering
  \includegraphics[width=8cm]{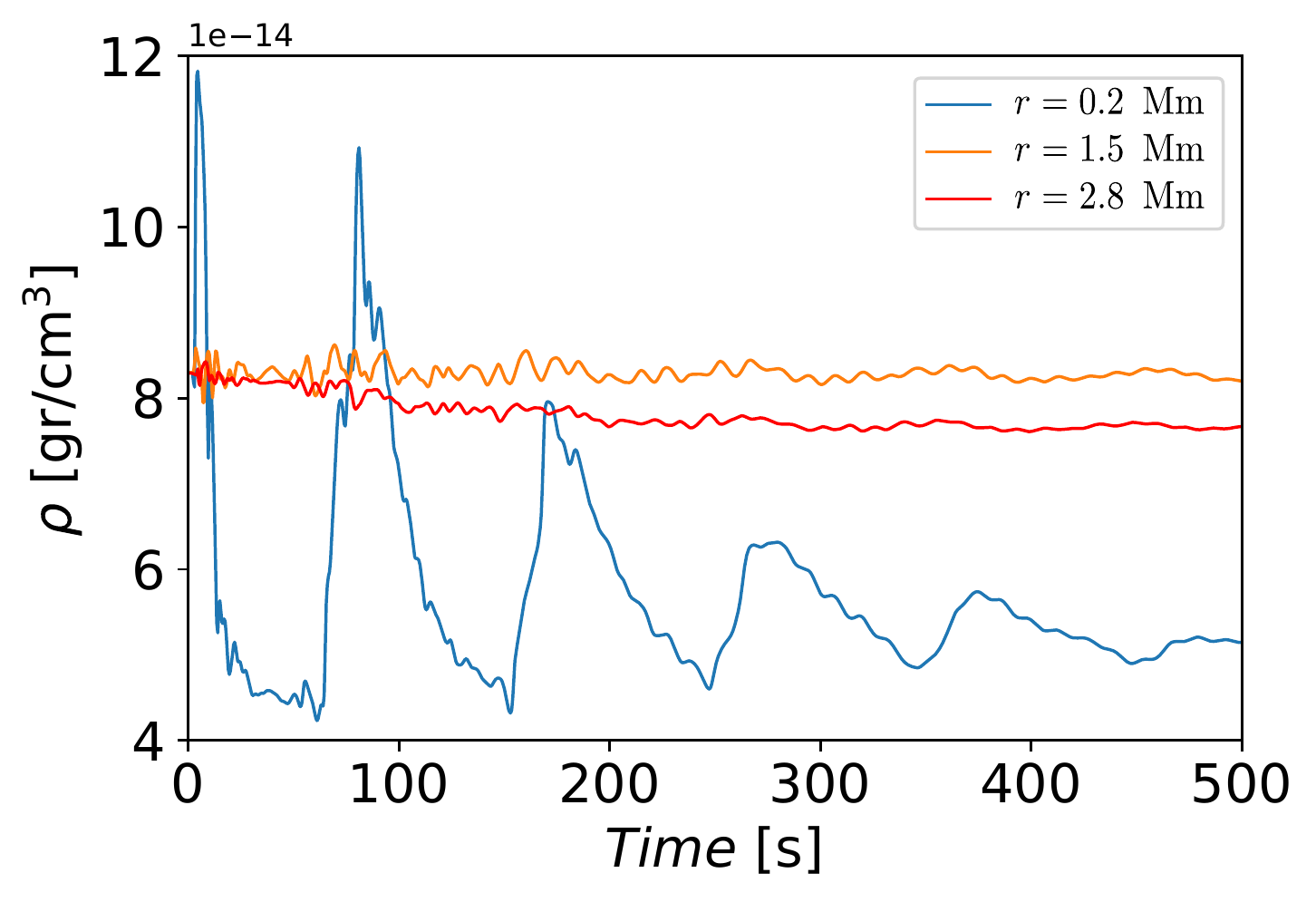}
  \caption{Density time evolution in three different points at the loop apex for the apex energy deposition.}
  \label{fig:pulsoapex_densapex}
 \end{figure}

 Figure~\ref{fig:pulsoapex_densapex} shows the density evolution at the mentioned three loop positions. 
 In the $z$-direction   two symmetric  slow shock wave fronts develop and travel in opposite directions channeled by the loop magnetic structure (also seen in Fig.~\ref{fig:shock_loop}). Behind each shock wave a rarefaction wave, travelling in the opposite direction and  emptying the medium, is excited. 
 The  density  behind the slow shock falls up to $\sim 4 \times 10^{-14}\,\text{g}\,\text{cm}^{-3}$ near the apex location (see  Fig.~\ref{fig:shock_loop}) (see also \citet{2009MNRAS.400.1821F}). 
 This is clearly seen for $p_1$ in Fig.~\ref{fig:pulsoapex_densapex}. 
 This shock pattern is damped at later times, i.e., due  to  successive encounters of the fronts with the chromosphere surface,   part of the shock energy is transferred to the dense chromospheric plasma.
 However, the density is weekly  perturbed for  the other point locations  ($p_2$ and $p_3$), because  the energy  deposition is concentrated around the loop axis due to the magnetic field collimation (see in Fig.~\ref{fig:shock_loop} that 
 points $p_2$ and $p_3$ are not reached by the vacuum region). The stationary 
  pattern established has a periodicity of $\sim 100\,\text{s}$. 

\begin{figure}[h!]
  \centering
  \includegraphics[width=9cm]{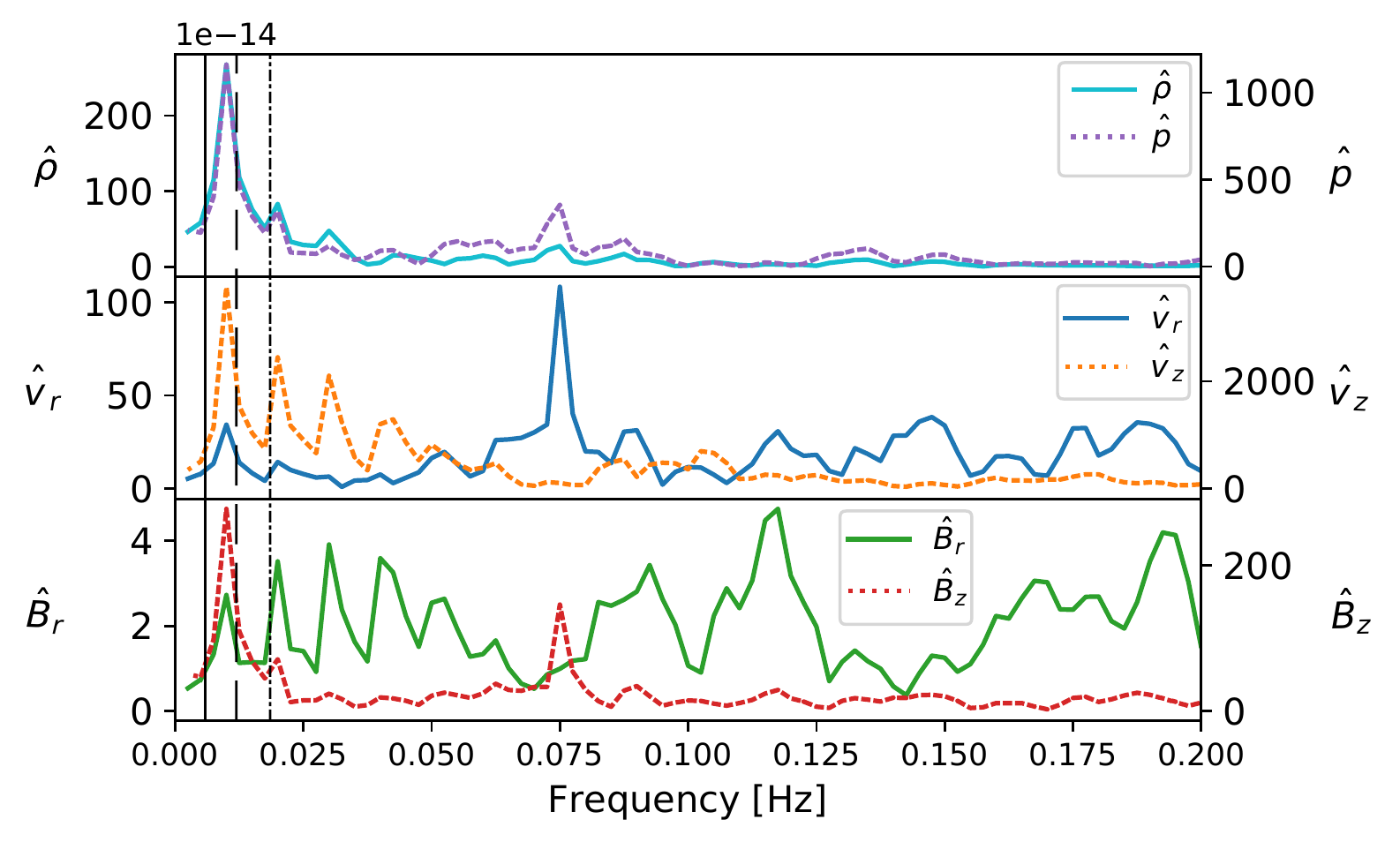}
  \caption{Fourier amplitude comparison of the different plasma parameters for the apex energy deposition measured at $p_1=(0.2,20)$Mm. The vertical lines correspond to the resonant frequencies: $\nu_{\text{res},1}$ (solid line), $\nu_{\text{res},2}$ (dashed line) and $\nu_{\text{res},3}$ (dot-dashed line).}
  \label{fig:pulsoapex_TFapex}
 \end{figure}

 To calculate the FT near the loop apex we choose to measure the significant $p_1$ signal and we evaluate it for the MHD variables:  density ($\hat{\rho}$), pressure ($\hat{p}$), radial velocity ($\hat{v}_r$), longitudinal velocity ($\hat{v}_z$), radial magnetic field ($\hat{B}_r$) and longitudinal magnetic field ($\hat{B}_z$).  For reference, the slow mode resonant cylinder frequencies  are: $\nu_{\text{res},1}=0.00492~$Hz, $\nu_{\text{res},2}=0.00984~$Hz and $\nu_{\text{res},3}=0.01476~$Hz ($\nu_{\text{res},n}=c_T n/2L$ where $n$ is an integer determining the parallel mode number, $L$ the cylinder's length, and $c_T$ the tube sound speed inside the cylinder). These frequencies are indicated with vertical lines in the FT  figures (Figs.~\ref{fig:pulsoapex_TFapex}, \ref{fig:pulsobase_TFapex}, \ref{fig:desbalance_TFapex}, \ref{fig:desbalance_TFbase}, \ref{fig:desbalance_energy_TFapex} and \ref{fig:desbalance_beta_TFapex}).
  Figure~\ref{fig:pulsoapex_TFapex} displays the $p_1$ FT for  the different  variables. Two main peaks are clearly distinguished considering all the variables: (I) $\nu = 0.01\,\text{Hz}$ (period time $\tau = 100\,\text{s}$) and (II) $\nu = 0.075\,\text{Hz}$ (period time $\tau = 13.3\,\text{s}$). The major peak of the density $\hat{\rho}$ occurs at frequency I.
The $z$-component of the velocity and the magnetic field present a peak at frequency I, which  is characteristic of slow modes. Comparing the shock speed ($\sim 379\,\text{km}\,\text{s}^{-1}$) with the local sound speed ($\sim 283\,\text{km}\,\text{s}^{-1}$) and the Alfv\'en speed ($\sim 499\,\text{km}\,\text{s}^{-1}$), we can see that this shock pattern corresponds to a slow magnetosonic shock.
 Also, the period is of the slow mode type ($\sim 100\,\text{s}$). 
 The $r$-component of the velocity has also a peak at the fast frequency II. However, the amplitude of $\hat{B}_z$ at this frequency is not intense enough to produce a fast mode.
  Also, the magnetic field $r$-component signal does not differ from the noise level. Thus, the whole picture suggests that the local pulse triggers a mode pattern of mainly two coupled frequencies that is strongly dominated by the slow one.

\subsubsection{At the footpoint}
\begin{figure}[h!]
 \centering
 \includegraphics[width=8cm]{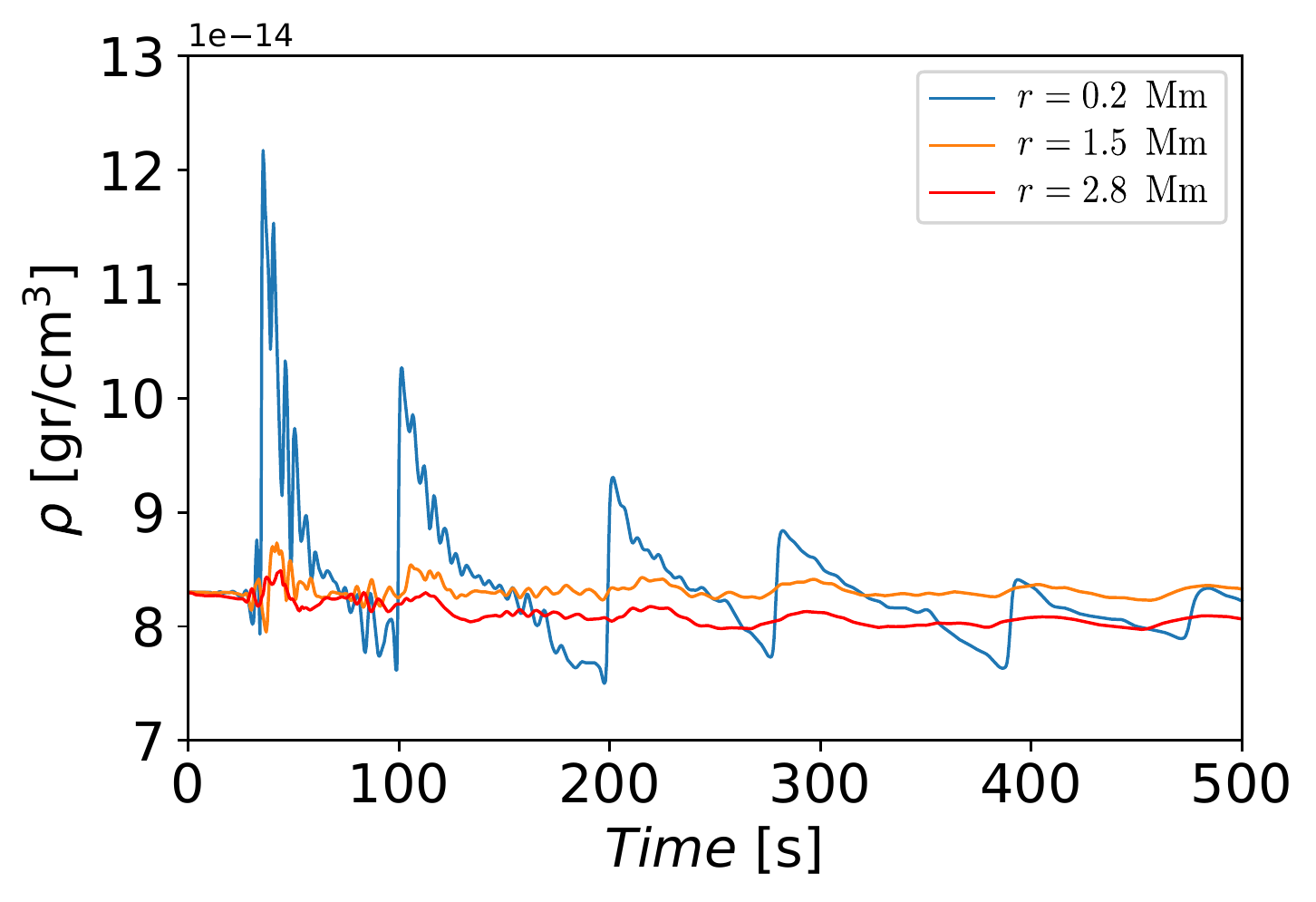}
 \caption{Density time evolution in three different points at the loop apex for the footpoint energy deposition.}
 \label{fig:pulsobase_densapex}
\end{figure}

 Figure~\ref{fig:pulsobase_densapex} shows the density evolution near the apex, i.e., for  $p_1$, $p_2$ and $p_3$, when  the pulse is located close to the chromospheric base of the loop at $r = 0\,\text{Mm}$, $z = 5.5\,\text{Mm}$. A similar shock pattern behaviour  is obtained with a time delay in the appearance of the shock front, i.e.  the shock front takes some time to reach the $p_i$ apex positions from the chromospheric place where the perturbation is triggered.
 As in the apex case a shock pattern is  measured for $p_1$. When the pulse source is located at the chromospheric base the rarefaction wave and the corresponding voided region do not initially perturb  the apex  and, therefore, the background density ($\sim 8.2 \times 10^{-14}\,\text{g}\,\text{cm}^{-3}$) is not modified.

\begin{figure}[h!]
 \centering
 \includegraphics[width=9cm]{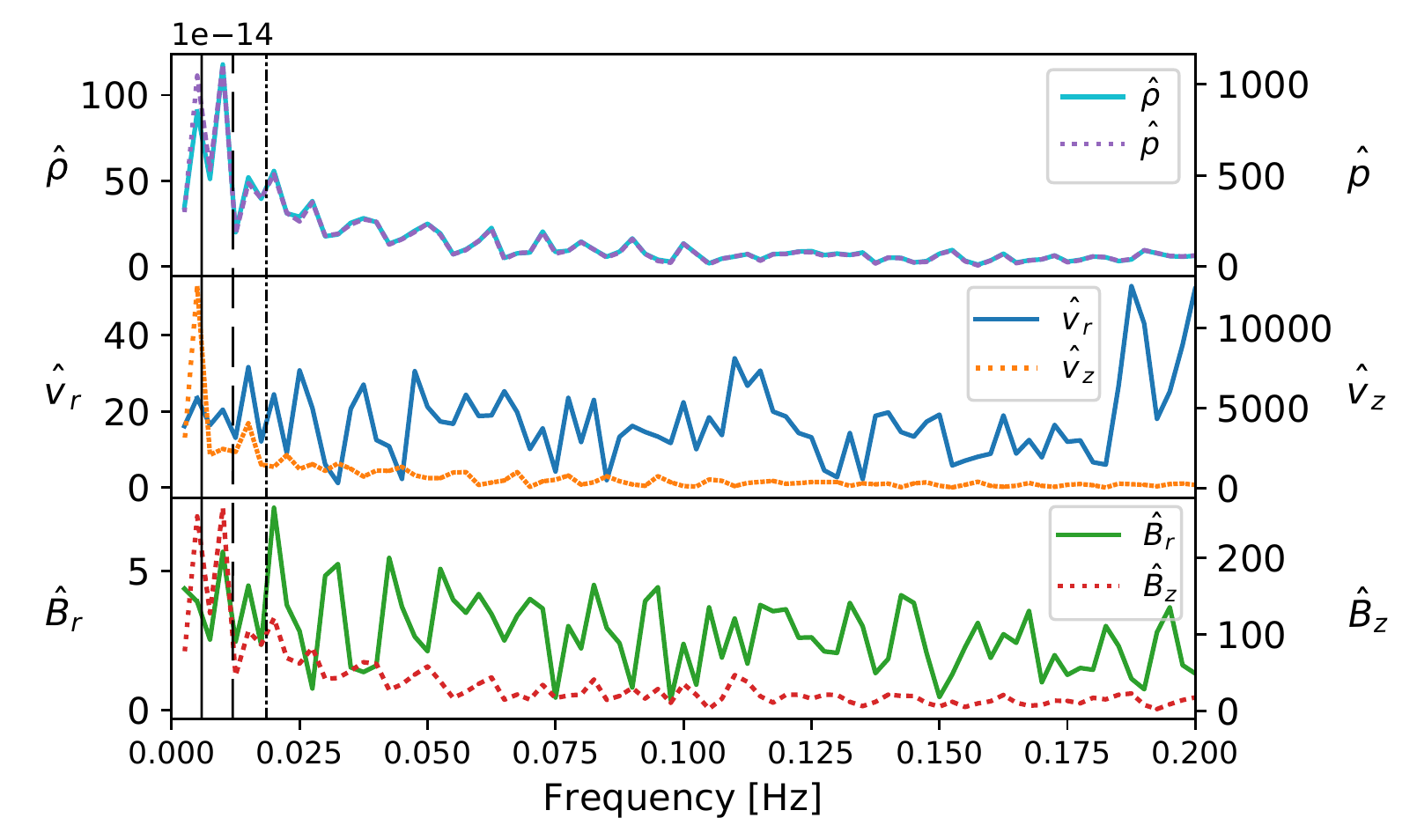}
 \caption{Fourier amplitude comparison for the footpoint energy deposition measured at $p_1=(0.2,20)$Mm. The vertical lines correspond to the resonant frequencies: $\nu_{\text{res},1}$ (solid line), $\nu_{\text{res},2}$ (dashed line) and $\nu_{\text{res},3}$ (dot-dashed line).}
 \label{fig:pulsobase_TFapex}
\end{figure}

 As before, Fig.~\ref{fig:pulsobase_TFapex} shows the FT for $p_1$, near the loop apex  for the density, pressure, velocity and magnetic field.
The pulse excites a main shock that travels during $100\,\text{s}$ from the chromospheric base into the coronal part of the loop towards the other loop end, where  it rebounds. 
 The repetition of this process, produces a cycle of positive (upwards) and negative (downwards) values of the $z$-component of the velocity every $200\,\text{s}$. The other shock, that travels into the  inner chromosphere, is mostly absorbed by these denser plasma layers and only a very small fraction of the wave can rebound.
  From the figure we  see a main peak of $\hat{v}_z$ at $\nu = 0.005\,\text{Hz}$, which is associated with this principal process. The density and the magnetic field intensity $B_z$ are enhanced every $100\,\text{s}$ (frequency I). The coupling with its first harmonics ($\nu = 0.005\,\text{Hz}$) are also seen in the respective  FTs.

 Note that the $r$-component of the velocity and magnetic field are not distinguished from a noise signal. 
 In this case, the whole picture suggests that the local pulse can only trigger a slow shock pattern due to its proximity to the chromosphere, where the line tied condition is fulfilled, i.e., the magnetic field tension is stronger and thus the $r$-component of the perturbation is relatively weak.

\mbox{}

To end this first analysis we mentioned that the increase of the pulse energy located in different positions of the loop (up to $3.7 \times 10^{28}\,\text{erg}$) derives in similar signal patterns where the enhanced fast component is --in almost the entire loop-- less important than the slow one. Thus, we find that a dominant fast mode cannot be obtain through a local deposition of energy.
In other words, a sausage mode is not possible to be developed when local energy depositions are used.
  
\subsection{Global instantaneous energy depositions} 
\begin{figure}[h!]
 \centering
 \includegraphics[width=9cm]{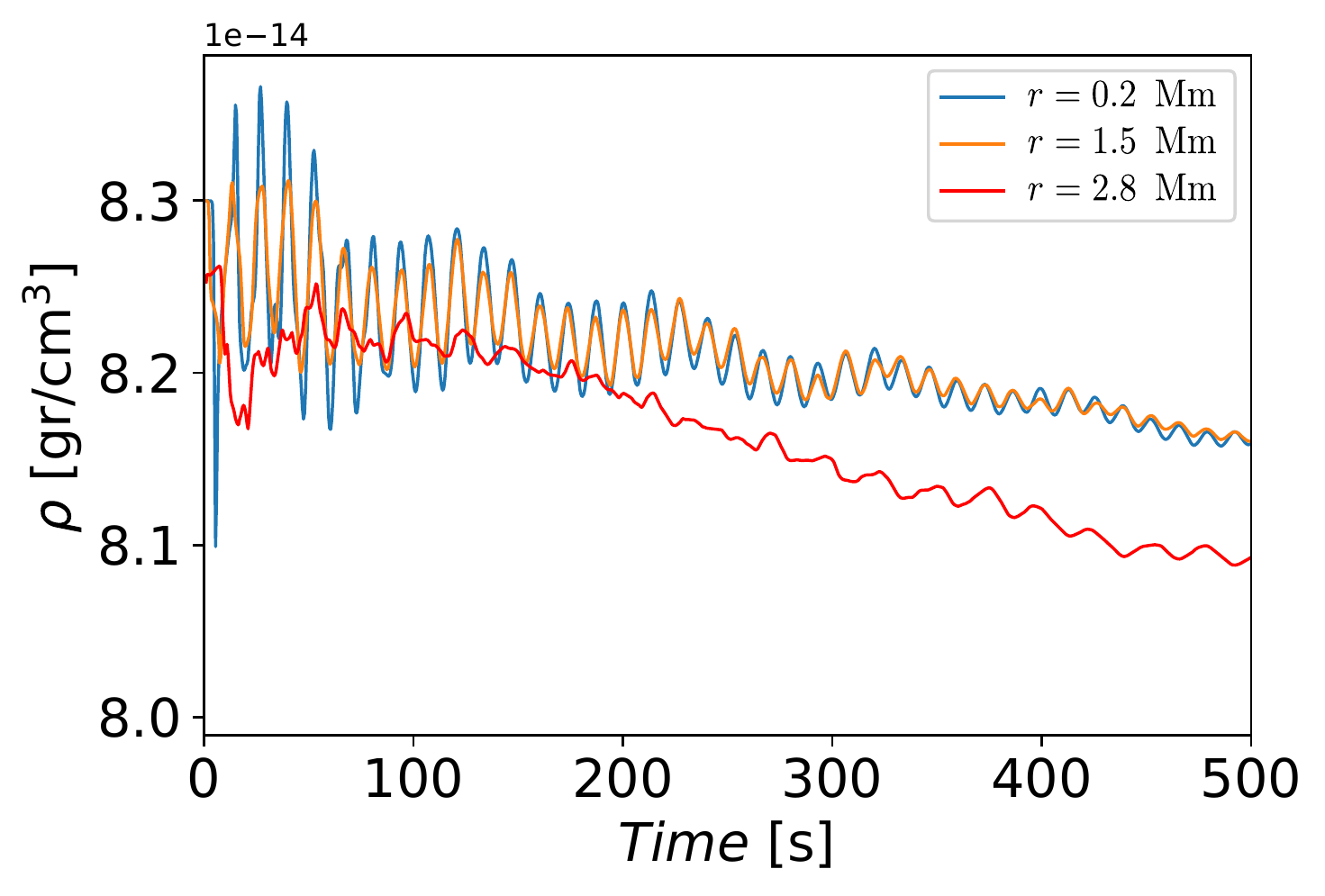}
 \caption{Density time evolution in three different points at the loop apex for the global energy deposition.}
 \label{fig:desbalance_densapex}
\end{figure}

As noted in  last section, using  a local deposition of energy resembling a typical microflare  (e.g., $1.5\times10^{27}\,\text{erg}$ and $3.7\times10^{28}\,\text{erg}$) it is not possible to generate a significant sausage mode perturbation.
In this section, taking into account the characteristic time considerations discussed in Section~\ref{perturbations}, we explore whether  sausage modes can be excited in the scenario of a global instantaneous energy deposition of the same amount than in the local case.    
Firstly, we inject $1.5\times10^{27}\,\text{erg}$ throughout the loop producing an imbalance between the loop and the corona. This extra energy is obtained by increasing the loop temperature up to $5.4\,\text{MK}$. The initial disequilibrium produces a shock wave that travels out of the loop, leaking part of its energy to the corona.
Then, the internal loop density starts to oscillate as can be seen in Fig.~\ref{fig:desbalance_densapex}. A new pattern of oscillations,  is obtained throughout the loop,  without shock waves and with small period ($13.3\,$s). The oscillation amplitudes at points $p_1$ and $p_2$ are much larger than those for $p_3$.
The perturbations  are modulated by a lower frequency and are damped while they evolve in time, due to the mentioned  leakage. The initial imbalance also produces a rarefaction wave that travels radially towards the loop axis, slightly decreasing its inner density. This is stabilized in about $1000\,\text{s}$.

\begin{figure}[h!]
 \centering
 \includegraphics[width=9cm]{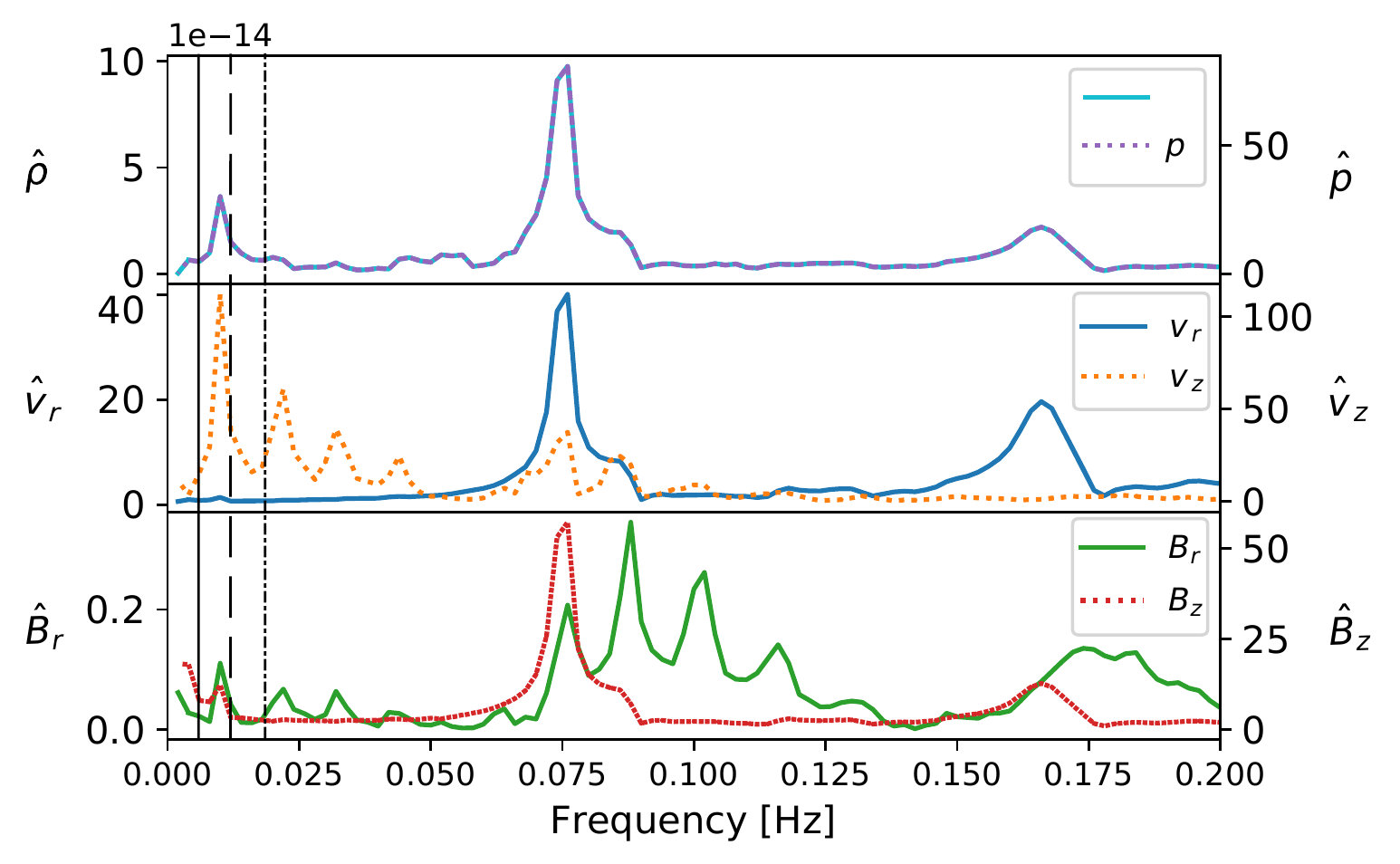}
 \caption{Fourier amplitude comparison for the ``imbalance''  at the apex, for the case with energy $E = 1.5\times10^{27}\,\text{erg}$ and parameter $\beta = 0.4$. The vertical lines correspond to the resonant frequencies: $\nu_{\text{res},1}$ (solid line), $\nu_{\text{res},2}$ (dashed line) and $\nu_{\text{res},3}$ (dot-dashed line).}
 \label{fig:desbalance_TFapex}
\end{figure}

Figure~\ref{fig:desbalance_TFapex} displays the FT for the density ($\hat{\rho}$), pressure ($\hat{p}$), radial velocity ($\hat{v}_r$), axial velocity ($\hat{v}_z$), radial magnetic field ($\hat{B}_r$) and axial magnetic field ($\hat{B}_z$) at $p_1$.
Contrary to the local deposition cases, in this global one, the main peak of the $\hat{\rho}$ FT signal occurs at the fast frequency II, and a much less intense peak occurs at the slow frequency I. 
Note that the fast frequency II is also the main peak for both, $\hat{v}_r$ and $\hat{B}_z$, corresponding  to a fast magnetosonic mode. Thus, the global instantaneous case, for the same energy content than the local cases, excites a coupled mode pattern mainly determined by a dominant fast magnetosonic mode and a weak slow mode, as can be seen, e.g., in the FT signal of $\hat{\rho}$, $\hat{v}_z$ and $\hat{B}_z$ at frequency I.

\begin{figure}[h!]
 \centering
 \includegraphics[width=9cm]{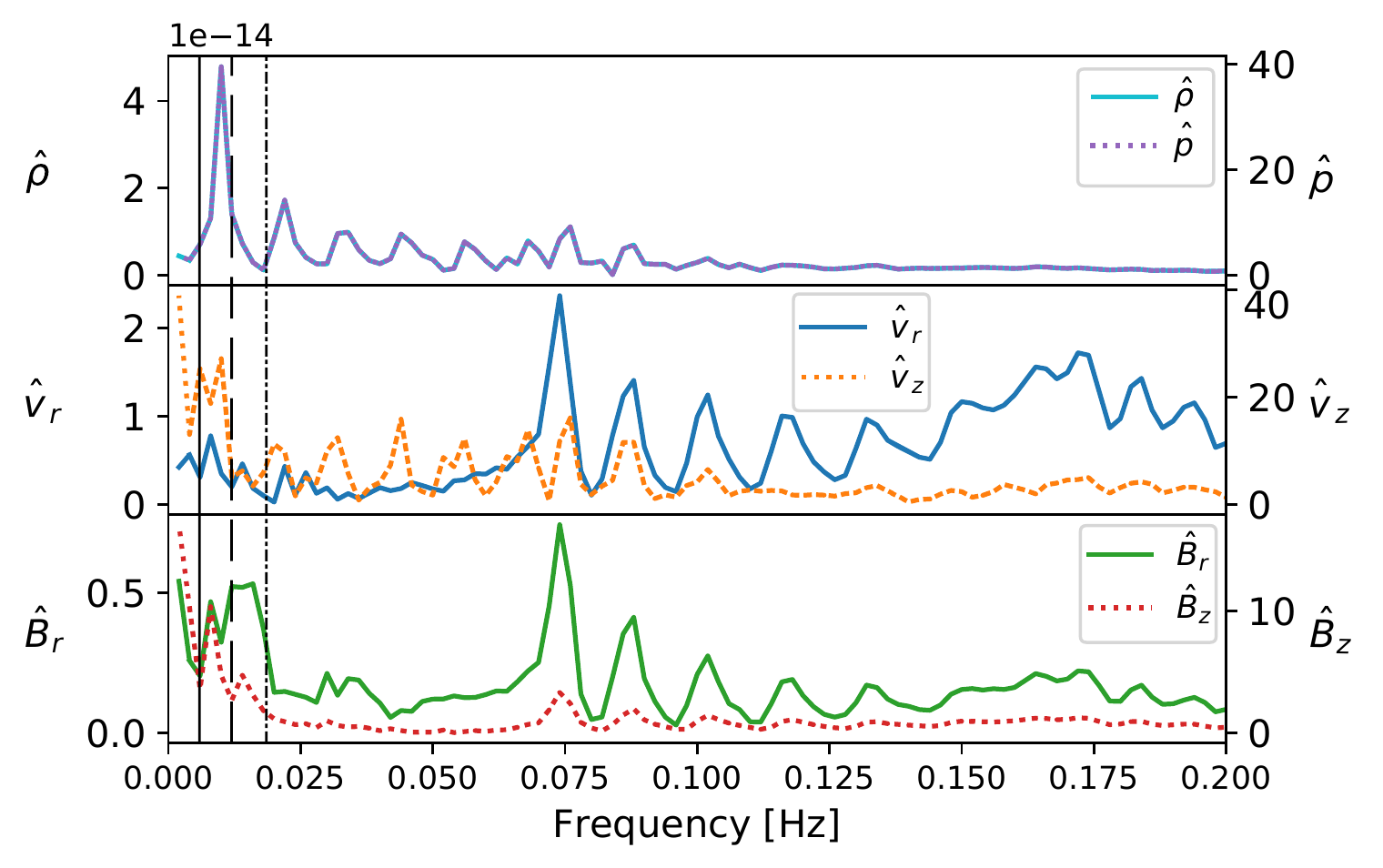}
 \caption{Fourier amplitude comparison for the ``imbalance''  at the foot. The vertical lines correspond to the resonant frequencies: $\nu_{\text{res},1}$ (solid line), $\nu_{\text{res},2}$ (dashed line) and $\nu_{\text{res},3}$ (dot-dashed line).
 }
 \label{fig:desbalance_TFbase}
\end{figure}

Figure~\ref{fig:desbalance_TFbase} is the same as Fig.~\ref{fig:desbalance_TFapex} measuring  the variables at the footpoint location $r = 0.2\,\text{Mm}$ and $z = 5.2\,\text{Mm}$. If we compare Fig.~\ref{fig:desbalance_TFapex} with  Fig.~\ref{fig:desbalance_TFbase} we see that, due to the line tied property, the fast mode vanishes at the chromospheric base \citep{2003A&A...412L...7N}. 

\subsection{The  importance of the $\beta$ parameter in the excitation of sausage modes}

The excitation of sausage modes requires the medium capability to regularly compress and restore the magnetic field lines. Hence, the thermal pressure must be comparable to the magnetic pressure and tension in order to balance the resistance of the magnetic field.  
As the $\beta$ parameter measures the relative importance between thermal and magnetic energy,  it seems an adequate parameter to analyse the capability of loop systems to excite sausage modes. We thus perform several numerical experiments to accomplish this task. Table~\ref{tab:table2} shows the $\beta$ parameter and energy values used for the different cases.
(With {\it reference case} we denote the global case of  previous subsection.) 

\begin{table}[h]
 \centering
 \begin{tabular}{l|c|c|c|}
  \cline{2-4}
     &  \multicolumn{3}{|c|}{Cases} \\
  \cline{2-4}
     & Reference     & Case I     &  Case II  \\
  \cline{1-4}
  \multicolumn{1}{ |c|  }{$T$ [MK]}   &   $3.1$  & $4.3$  & $4.1$   \\
  \cline{1-4}
  \multicolumn{1}{ |c|  }{$B$ [G]}   &      $51$     &    $45$      & $58$    \\
  \cline{1-4}
  \multicolumn{1}{ |c|  }{$\beta$}   &      $0.4$     &    $0.7$      & $0.4$    \\
  \cline{1-4}
  \multicolumn{1}{ |c|  }{Energy [$10^{27}\,$erg]}   &      $1.5$     &    $1.5$      & $37$    \\
  \hline
 \end{tabular}
 \caption{Temperature, magnetic field, $\beta$ parameter and energy values for the different cases.}
 \label{tab:table2}
\end{table}

\subsubsection{Case I: Enhanced $\beta$ parameter (same energy)}

\begin{figure}[h!]
 \centering
 \includegraphics[width=9cm]{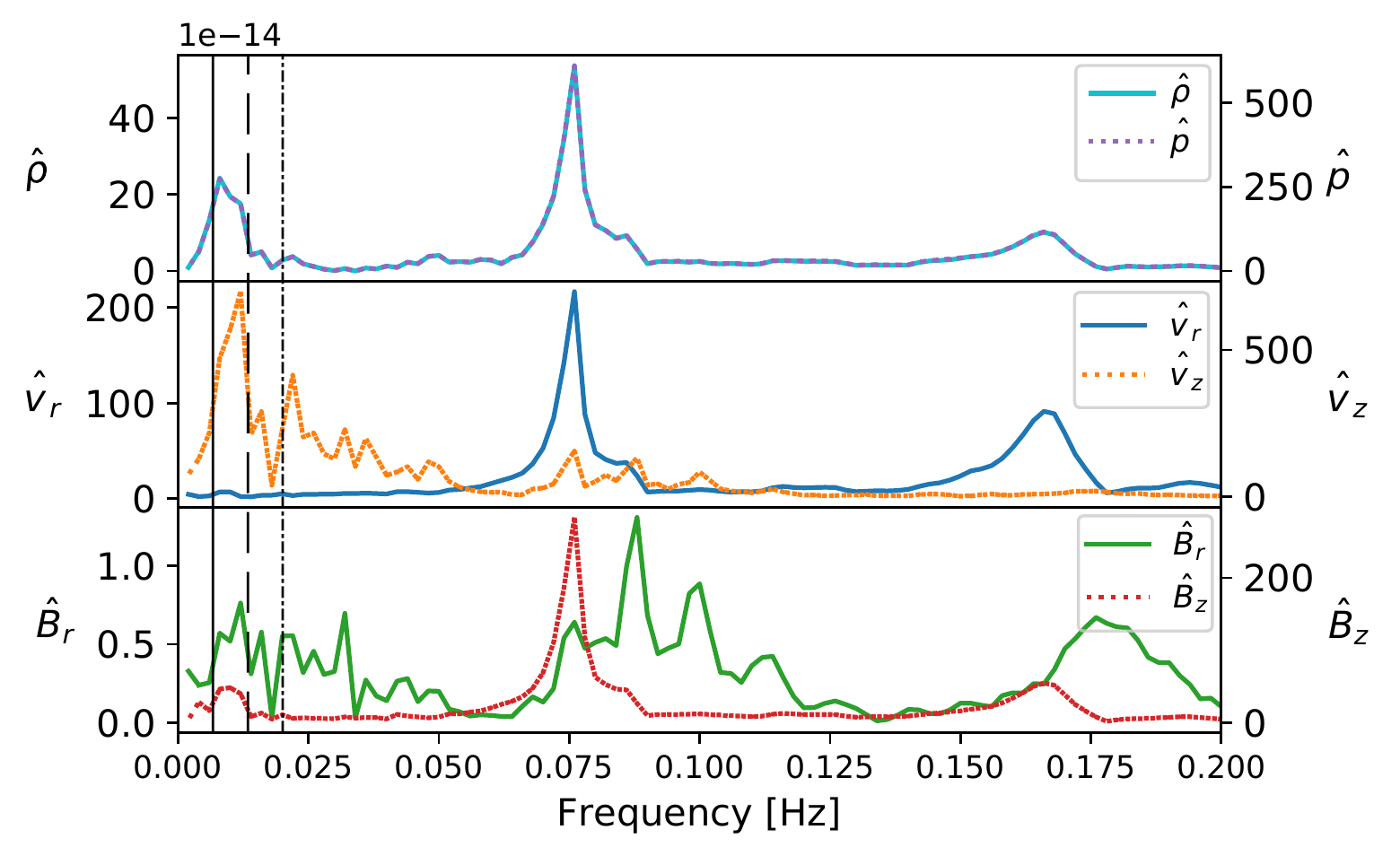}
 \caption{Fourier amplitude comparison for the ``imbalance''  at the apex, for the case with energy $E = 1.5\times10^{27}\,\text{erg}$ and  parameter $\beta = 0.7$. The vertical lines correspond to the resonant frequencies: $\nu_{\text{res},1}$ (solid line), $\nu_{\text{res},2}$ (dashed line) and $\nu_{\text{res},3}$ (dot-dashed line).
 }
 \label{fig:desbalance_energy_TFapex}
\end{figure}

In this experiment we increase the $\beta$ parameter from $0.4$ to $0.7$ keeping the same energy for the global perturbation than in the reference case ($E = 1.5 \times 10^{27}\,\text{erg}$).
Comparing Fig.~\ref{fig:desbalance_TFapex} with Fig.~\ref{fig:desbalance_energy_TFapex} we observe that an increase in the $\beta$ parameter leads to a notable increment in the FT amplitudes of  all the MHD variables. The FT amplitudes of the density, pressure, velocity and the longitudinal magnetic field component grow by a factor between $5$ and $8$ for both, the slow and the fast frequencies.
According to, e.g., \citet{1982SvAL....8..132Z}, this growth could be due to the increase in the temperature or  the decrease in the magnetic field (see eq. (6) of that paper).

\begin{figure}[h!]
 \centering
 \includegraphics[width=8.5cm]{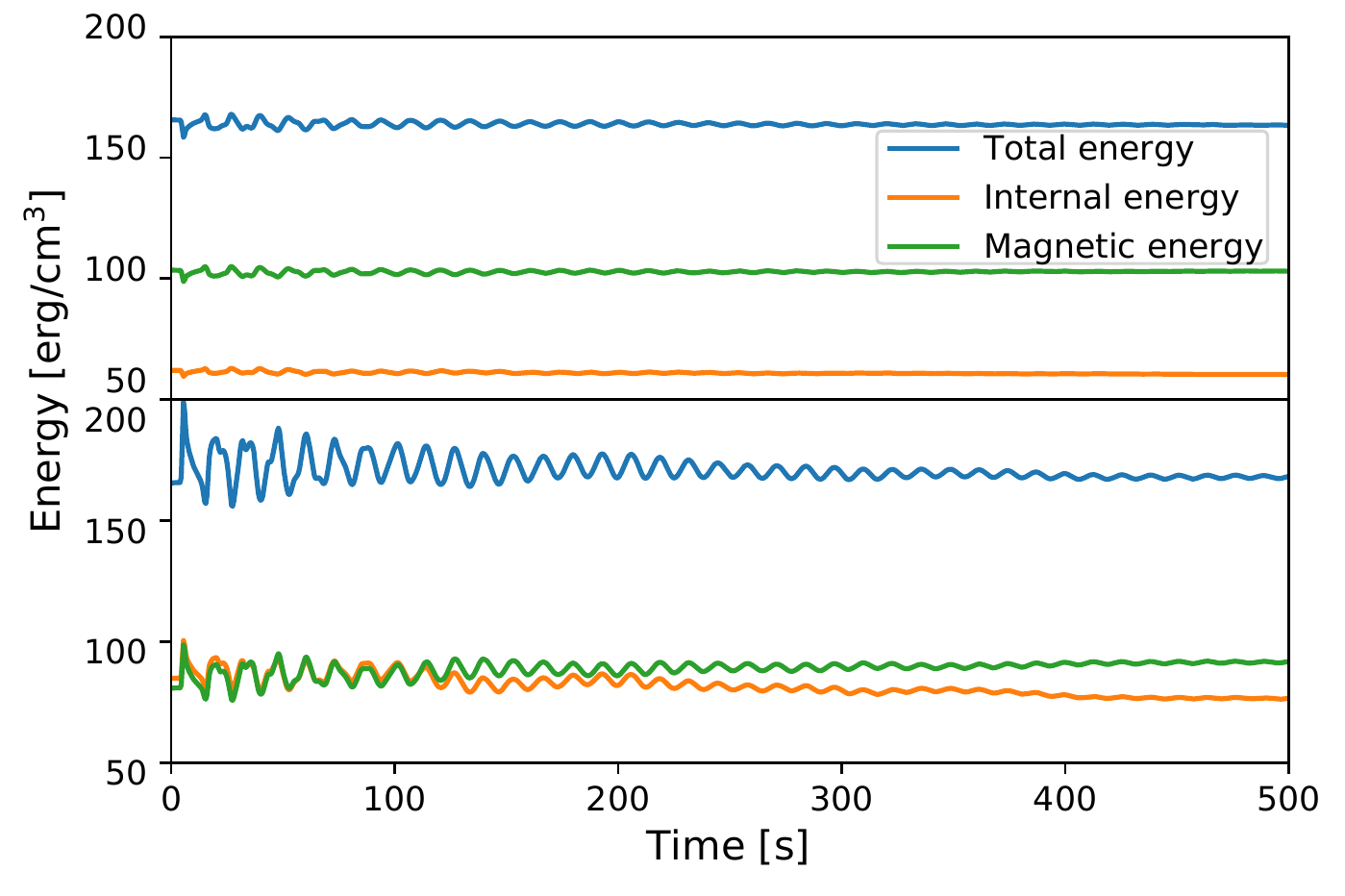}
 \caption{Evolution of the total, internal and magnetic energy for the ``reference case'' (upper panel) and ``case I'' (lower panel), measured at $p_1$.}
 \label{fig:energy}
\end{figure}

\begin{figure}[h!]
 \centering
 \includegraphics[width=8.5cm]{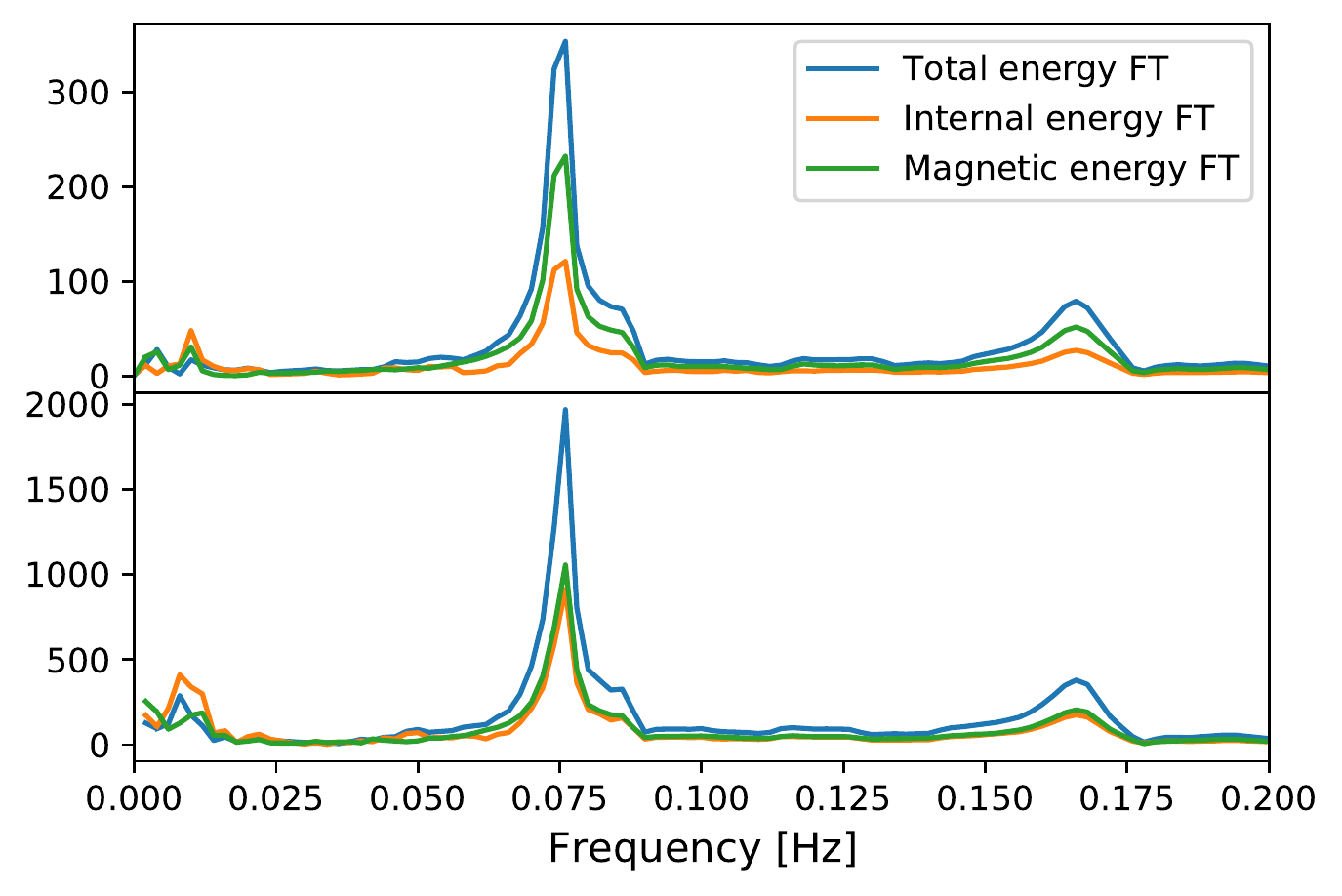}
 \caption{Evolution of the total, internal and magnetic energy Fourier amplitude for the ``reference case'' (upper panel) and ``case I'' (lower panel), measured at $p_1$.}
 \label{fig:TF_energy}
\end{figure}

To analyse the effect of the increase of $\beta$, in Fig.~\ref{fig:energy} we plot the evolution of the total energy (blue line), the magnetic energy (green line) and the internal energy (orange line) for both $\beta$ cases at $p_1$. Note that the amplitude of the oscillation is increased when $\beta$ is augmented and it oscillates around the initial total energy value. Thus, different $\beta$ values lead to different amplitude responses. 
As the reference case and Case I have almost the same slow and fast frequencies, with a difference of $4\%$, it seems that the major amplitude response of Case I is due to the smaller magnetic field value --which is  associated with the less rigid behaviour of the magnetic field lines--,  and also to the larger temperature value --leading to a larger capability to bend the magnetic field lines of the loop system. 
 The oscillatory pattern and the energy distribution between modes can also be analysed in detail from the FTs shown in Fig.~\ref{fig:TF_energy}.
As seen in the figure, a similar behaviour is observed, i.e., the FT amplitudes of the slow and the fast frequencies for the total, magnetic and internal energies strongly increase when the $\beta$ parameter is augmented.
In addition, we can see that, as expected, the slow mode energy content is mainly due to the contribution of the internal energy, almost independently of the value of the $\beta$ parameter, while the energy distribution of the fast mode directly depends on the initial relation between the magnetic and the internal energy, i.e.,  depends directly on the $\beta$ parameter.
As a final result, we can say that this experiment shows that the raise of the $\beta$ parameter in the global case strongly increases the FT amplitudes of  all the variables for both, the slow and the fast modes. Hence while $\beta$ approaches one, i.e., in coronal regions of relatively high values of $\beta$, the sausage modes are more easily excited.

To compare the relative importance between the fast and the slow components in Table~\ref{tab:table3} we show the rate between the fast FT peak and the slow FT peak of the density for each case. For the reference case, the rate between the fast and slow mode is $2.7$, which shows a dominance of the fast mode. When the $\beta$ parameter is larger, this rate is almost the same: $2.2$.
To obtain a larger $\beta$ parameter with the same initial energy we increase the temperature and decrease the magnetic field. The dynamics results more hydrodynamic favouring the slow mode. Hence, this explains the lower rate obtained in relation with the reference case seen in Table~\ref{tab:table3}.

\begin{table}[h]
 \centering
 \begin{tabular}{l|c|c|c|}
  \cline{2-4}
     &  \multicolumn{3}{|c|}{Cases} \\
  \cline{2-4}
     & Reference     & Case I     &  Case II  \\
  \cline{1-4}
  \multicolumn{1}{ |c|  }{Rate }   &  $2.7$   & $2.2$  & $5.0$    \\
  \hline
 \end{tabular}
 \caption{Rate between the fast FT intensity and the slow FT intensity for the density variable.}
 \label{tab:table3}
\end{table}

\subsubsection{Case II: Enhanced energy (same $\beta$ parameter)}
\begin{figure}[h!]
 \centering
 \includegraphics[width=9cm]{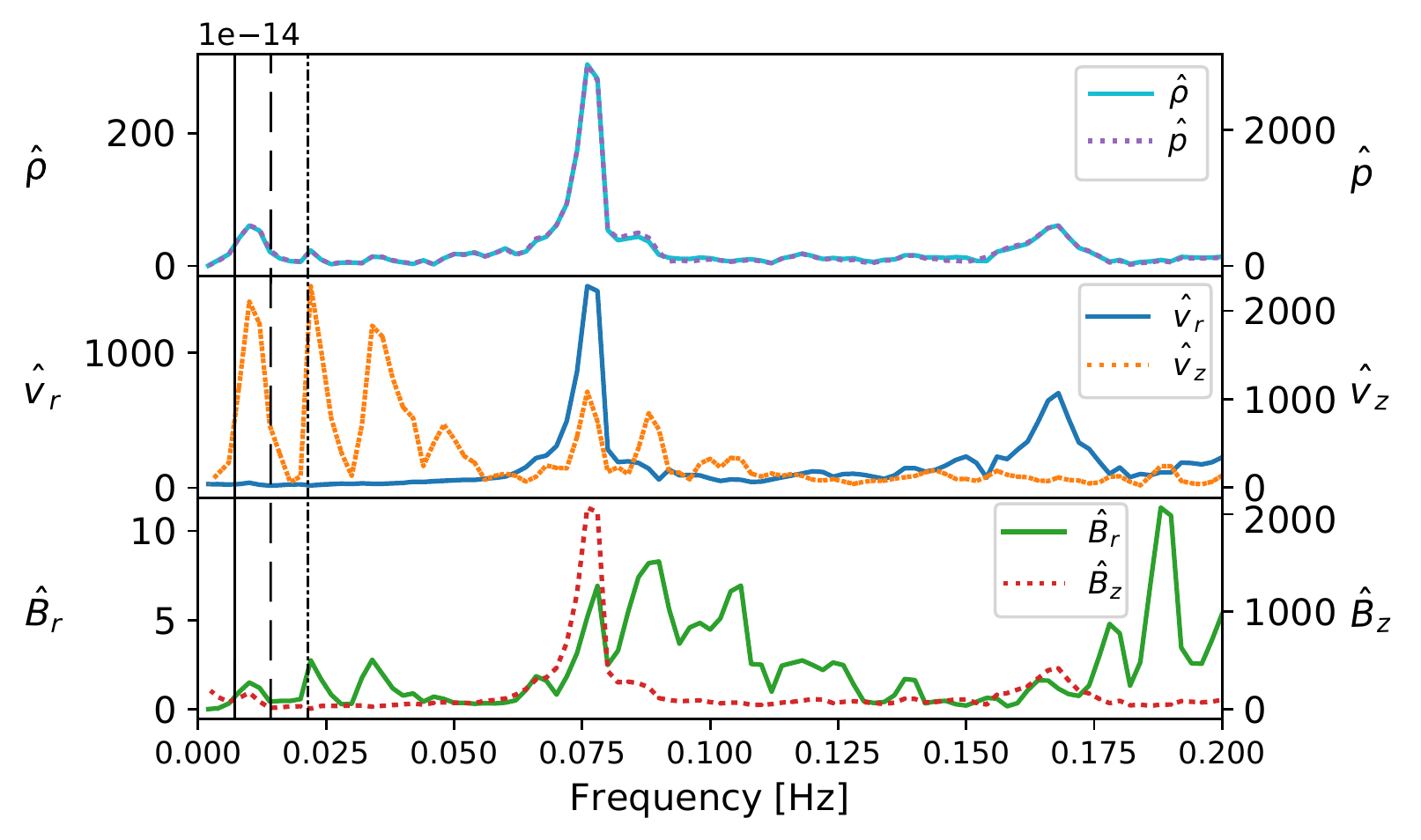}
 \caption{Fourier amplitude comparison for the ''imbalance''   at the apex, for the case with energy $E = 3.7\times10^{28}\,\text{erg}$ and parameter $\beta = 0.4$. The vertical lines correspond to the resonant frequencies: $\nu_{\text{res},1}$ (solid line), $\nu_{\text{res},2}$ (dashed line) and $\nu_{\text{res},3}$ (dot-dashed line).}
 \label{fig:desbalance_beta_TFapex}
\end{figure}

In this experiment the energy is increased, from $1.5\times10^{27}\,\text{erg}$ to $3.7 \times 10^{28}\,\text{erg}$, keeping the same $\beta$ parameter as in the reference case ($\beta = 0.4$), see~Table~\ref{tab:table2}.
Comparing Fig.~\ref{fig:desbalance_TFapex} with Fig.~\ref{fig:desbalance_beta_TFapex} we see a notable increase in the FT components of all variables (at least by a factor of $10$). 
The overall picture implies that the change in energy with $\beta$ constant changes the proportionality between the fast and the slow FT contributions (see Table~\ref{tab:table3}). 
When the energy increases one order of magnitude, the rate between fast and the slow modes is twice the rate of the reference case.
   
\mbox{ }

Summarizing, the numerical experiments given in Table~\ref{tab:table3} considering a global deposition of energy that resembles the action of typical microflares, indicate that these perturbations lead to mode patterns where the fast sausage mode is dominant in almost the entire loop\footnote{We also performed experiments increasing the $\beta$ parameter of the loop and the energy pulse in the local deposition cases. 
These experiments did not lead to a dominant sausage mode as those shown in Section~\ref{sec:local}. }.

\section{Conclusions}

In this work, we studied the oscillatory modes that can be excited by  microflare depositions of energy in the interior of coronal loops. Based on the evidence that the elementary bursts seem to arise from a single flaring loop than from various of them, we designed a set of numerical experiments in which the motion is triggered by an instantaneous deposition of energy.
Considering different relations between the thermal conduction time and the radiative cooling time, we defined two situations that could excite an oscillatory pattern in the loop interior, a ``local deposition of energy'' when the radiative cooling effect is dominant and a ``global deposition of energy'' when the heat conduction prevails.

In the local energy deposition analysis we found that, for energy levels of the order of typical microflares, the local pulse triggers a mode pattern of  mainly  two coupled frequencies which is strongly dominated by the slowest one.
A pair of two opposite slow shocks are detected travelling along the magnetic field lines due to the strong collimation until reaching the chromospheric surface, where the shocks rebound producing the oscillatory motion.
This is in agreement with \citet{2009MNRAS.400.1821F}, who suggested that to detect an internal loop slow mode a density contrast produced by a shock wave is required.
This shock pattern, strongly collimated along the loop axis, is the main feature (of typical slow speeds and slow periods) that characterizes every local energy deposition where fast mode signatures are  weak enough, i.e., sausage modes are not  detected under this scenario.

In the analysis of the global energy deposition we showed that a pattern of coupled modes is obtained, where the fast magnetosonic one is dominant, i.e., the sausage mode is feasible to be developed for a global deposition of energy.
These results suggest that the rarity of fast sausage modes is due to the requirement that the heat conduction effect must be dominant in this model.
Such a requirement is almost exclusively achieved in active flaring regions where the temperature is high enough and the loop length is short.

Knowing that global energy deposition can produce sausage modes, we explore the influence of the $\beta$ parameter and the effect of the energy level in the strength of the fast signal.
For low $\beta$ parameters the magnetic forces dominate the plasma dynamics and the magnetic field resists the compression generated, for instance, by fast transverse modes, as it happens in the corona where the sausage modes are rarely observed.
For larger $\beta$ parameters, which can occur when the thermal energy overcomes the magnetic field resistance, we found that the fast signal is increased almost in the same amount as the slow one.
On the other hand, the increment of the energy level keeping constant the $\beta$ parameter produces a substantial rise of the fast signal over the increase in the slow one.
These effects are not detected when either the $\beta$ parameter or the energy level are increased in the local energy deposition scenario, which reinforces the hypothesis that a global energy deposition is needed to obtain a sausage mode.

With respect to the strong damping of sausage modes it is clear that, having used an ideal MHD model, this process is not associated with dissipation mechanisms.
This is in agreement with the fact that the corona is a highly conductive medium where the resistivity can be neglected.
Our  experiments showed that the damping of modes is due to the coupling and the transfer of energy between modes, the absorption at the chromospheric bases and the leakage across the boundaries of the loops.

\begin{acknowledgements}
We thank the anonymous referee for very useful comments that helped us to improve the previous version of this manuscript. HC is doctoral fellow of CONICET. MC, AC, GK and OR are members of the Carrera del Investigador Cient\'ifico (CONICET). HC, MC y GK acknowledge support from ANPCyT under grant number PICT No. 2016-2480. MC also acknowledge support by SECYT-UNC grant number PC No. 33620180101147CB. Also, we thank the Centro de C\'omputo de Alto Desempe\~no (UNC), where the simulations were carried out.
\end{acknowledgements}

%
   \bibliographystyle{aa} 
   \bibliography{biblio} 
%

\end{document}